\documentclass[12pt]{article}
\usepackage{xcolor}
\usepackage{tikz}
\usepackage{pgfplots}
\pgfplotsset{compat=newest}
\begin{document}
\begin{titlepage}
\title{ Modified Newtonian dynamics and excited hadrons}
\author{
{Piotr \.Zenczykowski }\footnote{E-mail: piotr.zenczykowski@ifj.edu.pl}\\
{\em Professor Emeritus}\\
{\em The Henryk Niewodnicza\'nski Institute of Nuclear Physics}\\
{\em Polish Academy of Sciences}\\
{\em Radzikowskiego 152,
31-342 Krak\'ow, Poland}\\
}
\maketitle
\begin{abstract}
We discuss implications
 that result from
the acceptance of MOND parameter $a_M$ as the third fundamental
constant of nature in addition to $c$ and $h$. 
To this aim we use the concept of Newtonian triangle
in the $(log(r),~ log(m))$ plane and show that excited states on
hadronic Regge trajectories
lead into the MOND regime beyond this triangle. The experimentally suggested
freezing of one internal spatial degree of quark freedom in excited baryons
corroborates then the conjecture that MOND is related to  the idea of space emergence. 
\\
\end{abstract}

\vfill
\end{titlepage}

\section{Introduction}
Among many dimensional constants of nature one may distinguish those that are thought to apply universally and thus are deemed more fundamental than others. Three such constants
are widely believed to be relevant in setting the `natural' units of mass, space, and time. These are the speed of light $c$, Newton's gravitational constant $G$, and Planck's constant $h$. They are thought to define the scales of spaciotemporal and quantum  properties of nature.
Using $c=3 \times 10^{10} cm/s$, $G=6.67 \times 10^{-8} cm^3/(g~s^2)$, and $h=6.62\times 10^{-27} g~ cm^2/s$, dimensional considerations define then the corresponding distance scale, ie. 
Planck's length 
\begin{equation} 
\label{lP}
l_P=\sqrt{hG/c^3}=4.05 \times 10^{-33}~cm,
\end{equation}
 at which transition from the quantum to the classical description of space is generally believed to occur.
    With the corresponding Planck mass coming out surprisingly large, ie.
   \begin{equation}
   \label{mP}
   m_P=\sqrt{hc/G}=5.46\times 10^{-5}~g,
   \end{equation}
   Planck's length is often considered as a scale relevant for space alone, ie. independent of matter mass scales.\\
   
Now, as demonstrated by Meschini \cite{Meschini}, dimensional analysis is not a trustworthy tool as it leads to physically correct conclusions only when the underlying theory
  is already known. Furthermore, on general philosophical grounds, no theory of ours can be valid everywhere: a set of universal physical concepts should be considered adequate only within some limited  domain. After all, even the best of our `universal' theories constitute a {\it description} of reality, which description is - consequently - only approximate 
 and idealized.\footnote{\label{ff} Confusion of theory with reality is a common case of
 the fallacy of misplaced concreteness (reification) \cite{Whitehead}.
 In the words of Heisenberg \cite{Heisenberg} ``...the scientific concepts are idealizations. (...) through this process of idealization and precise definition the immediate connection with reality is lost. The concepts still correspond very closely to reality in that part of nature which had been the object of the research. But  the correspondence may be lost in other parts containing other groups of phenomena.''
 } Therefore, an important question is how far the universality in question may be extended.\\
 
 Now, with respect to the question of fundamental constants, there seems to be no deep reason why one should not consider the cosmological constant 
$\Lambda = 1.19 \times 10^{-56} cm^{-2}$ (which enters Einstein field equations alongside $G$) as just as relevant in setting the distance and mass scales as $G$, $c$ and $h$.
 Implications of the introduction of $\Lambda$ were discussed at length in \cite{Zen2018}, where dimensional analysis was used to support the thesis that one should be able to learn about the nature of space by studying properties of elementary particles at hadronic mass and distance scales. 
{The arguments used in \cite{Zen2018} involved also 
the generalization of the concept of mass as suggested by the phase-space-based explanation 
\cite{Zen2008,ZenBook,Clifford}
of the Harari-Shupe rishon model \cite{HSmodel} of elementary particles, and the phenomenological conclusions on the internal spatial structure of excited baryons
\cite{Capstick}.} In the following these issues will be briefly examined from a somewhat different angle.

\section{The Newtonian Triangle}
In view of the vulnerability of dimensional analysis \cite{Meschini} it would be advantageous to replace it with some more physical argumentation. Now, 
the three constants $c$, $G$, and $h$ are of two different types. As noted by Milgrom \cite{Milgrom2019},
$G$ works as a conversion factor  that links the concepts of gravitational and inertial
masses. On the other hand, $c$ and $h$ impose physically defined bounds on the applicability regime
corresponding to classical Newtonian dynamics,
specifying in particular the lower bound ($l_P$) on the applicability of the classical concept of distance.
In order to bound it
from above in the following we shall replace $\Lambda$ with a  universal constant physically akin to $c$ and $h$ (rather than to $G$). Such a constant must exist\footnote{Infinite extrapolations are not admissible in physics. 
} and, as it turns out, there is a good candidate for it. It was originally proposed by Milgrom \cite{Milgrom1983} as a way to solve some specific problems identified in the astrophysical realm. The idea was that beside constants $h$ and $c$ that limit the sizes of action and velocity, there exists another constant ($a_M$) that provides a lower bound on the acceleration in the standard Newtonian description of gravity-induced motions. As shown below, adding $a_M$ to $h$ and $c$ completes the procedure of specifying  the boundary of the phase-space (distance-mass) domain where classical Newtonian dynamics may be considered applicable.\\

Originally, the existence of $a_M$ was proposed as a way to solve the problem of the unexpectedly
fast motion of stars in the outer reaches of 
spiral galaxies. 
Two conceptually different solutions of the problem may be formulated. The first (and more popular) paradigm accepts standard Newtonian dynamics and 
explains the effect through the introduction of additional, nonluminous, and so far undetected ``dark matter" that builds huge halos in and around the galactic structures. The other paradigm proposes that in this situation the boundary of the applicability of Newtonian physics has already been reached. Specifically, for acceleration $a$ smaller than $a_M=1.2 \times 10^{-8}~cm/s^2$ (as determined from astrophysical data)
the Newtonian expression for the acceleration induced by gravitational source of mass $m$, ie. $a=a_N\equiv Gm/r^2$, is supposed to be modified to
\begin{equation}
\label{MOND1}
a = \frac{\sqrt{Gm}}{r}\sqrt{a_M},
\end{equation}
thus defining the basic ingredient of the MOND (Modified Newtonian dynamics) approach, 
a growing in strength competitor of the dark matter paradigm 
\cite{McGaugh2014}.\footnote{
In \cite{MendozaReview} the contemporary idea of dark matter is discussed alongside two 19th century cases of `dark matter' explanations of small non-Newtonian anomalies observed in the motions of Uranus and Mercury. It is stressed in \cite{MendozaReview} that, irrespectively of whether the correct explanation  was the existence of `dark matter' (ie. Neptune in the case of Uranus) or a modification of theory (General Relativity in the case of Mercury), the similarity with the modern dark matter explanations is quite remote,
as the latter, unlike the 19th century cases, requires huge energy corrections. Consequently, 
the current case of dark matter seems more similar to another 19th century concept: that of luminiferous aether, the existence of which turned out to be the figment of physicists' imagination, driven by a misplaced idea of concreteness.}
 Comparing Eq. (\ref{MOND1}) with $a_N$ we see that for the source of mass $m$ the transition to MONDian dynamics occurs around radius
\begin{equation}
\label{rM}
r_M(m) = \sqrt{\frac{Gm}{a_M}}.
\end{equation}
Note that the radius $r_M(m)$ above which Newtonian dynamics gives way to the MONDian
description depends on the size of the field-generating mass $m$.\\

Wide applicability of the basic MOND assumption (\ref{rM}) 
is corroborated by observations and analyses over  12 orders of magnitude in the mass of astrophysical structures considered, to name only galaxy clusters, spiral galaxies, dwarf spheroidal galaxies, and as far down in mass as wide binaries (see Fig. 1 in \cite{Hernandez}).
Furthermore, there are intriguing order-of-magnitude coincidences between  astrophysically and cosmologically defined acceleration parameters \cite{Milgrom2020}, namely
$cH_0 \approx c^2\sqrt{\Lambda}\approx a_M$
(with $H_0$ and $\Lambda$ being the Hubble and cosmological `constants'),
which extend phenomenological usefulness of $a_M$ to cosmological scales, ten orders of magnitude above galaxy clusters. 
The large span of the applicability of $a_M$ strongly suggests its fundamental nature. Such a point of view was discussed in many papers see eg.  \cite{Milgrom2020,MendozaReview,Capozziello2011,ExtendedMetricGravity}.
For an extensive presentation of the pro-MOND astrophysical 
arguments see eg. \cite{McGaugh}. 
{To the  list given in \cite{McGaugh} one should add the case of wide binaries \cite{WideBinaries} and the most recent identification of the external field effect in galactic structures \cite{EFE2020}. The latter
observation supplied MOND ideas with a strong boost
as the effect is unique to MOND and does not appear in Newton-Einstein gravity.} Thus, the universality of the MOND bound on acceleration at astrophysical scales  may be  
assumed.\\

Just as in the case of Newtonian/MONDian boundary (\ref{rM}), analogous mass-dependent radii specify the boundaries between the domain of classical Newtonian dynamics and the relativistic regime or the quantum region.
The transition between the relativistic (black-hole) and Newtonian domains occurs at the distance scale defined by the Schwarzschild radius 
\begin{equation}
\label{rS}
r_S(m) = 2Gm/c^2,
\end{equation}
which depends on the black hole mass $m$ (with the Newtonian description applicable for $r>>r_S(m)$). \\

Similarly, to estimate the distance scale characteristic for the quantum-classical transition  
  associated with mass $m$ one may use
the universal concept of the relevant Compton wavelength:
\begin{equation}
\label{Compton}
r_C(m)=\frac{h}{mc},
\end{equation}
with the classical description applicable for $r>>r_C(m)$.\\

On the ($log(r),log(m)$) phase-space plane considered in \cite{Hernandez} (see Fig. 1) the three functions (\ref{rM},\ref{rS},\ref{Compton}) define the sides of a triangle  (the MONDian side, the black hole side, and the quantum side - marked in Fig. 1 respectively by: dashed, solid, and solid lines). The interior of this triangle (except for the transition regions close to its borders)  constitutes the domain  where classical Newtonian dynamics applies (denoted by $N$).  The vertices of this `Newtonian triangle' are located at the points defined by the
following values of $m$ and $r$:\\

1) Planck (Compton-Schwarzschild):
\begin{equation}
\label{mCS}
m_{CS}=\sqrt{\frac{hc}{2G}} \approx m_P,
\end{equation}
\begin{equation}
\label{rCS}
r_{CS} = \sqrt{\frac{2hG}{c^3}}\approx l_P,
\end{equation}

2) Universe (MOND-Schwarzschild):
\begin{equation}
\label{mMS}
m_{MS}=\frac{c^4}{4Ga_M} 
\approx 2.5\times 10^{56}~g \approx m_U,
\end{equation}
\begin{equation}
\label{rMS}
r_{MS}= \frac{c^2}{2a_M} = 3.8\times 10^{28}cm \approx r_U,
\end{equation}

3) Hadron (Compton-MOND):
\begin{equation}
\label{mCM}
m_{CM}=\left(\frac{h^2a_M}{Gc^2}\right)^{1/3} = 0.2 \times 10^{-24}g \approx m_H,
\end{equation}
\begin{equation}
\label{rCM}
r_{CM}=\left(\frac{Gh}{a_Mc}\right)^{1/3}=10^{-12} cm \approx r_H.
\end{equation}
Planck's mass and length appear in (\ref{mCS}) and (\ref{rCS}) as the lower (quantum) bound on the size
of a black hole (see Fig. 1). Given its derivation from black-hole properties (Eq. (\ref{rS})), 
$l_P$ should not be viewed as a mass-independent
distance scale emerging from the quantum substratum, as it is
quite often imagined in various emergent space approaches.
Similarly, the scales of Universe's mass $m_U$ and size $r_U$ appear in (\ref{mMS},\ref{rMS}) as the upper (MONDian) bounds
on the size of a black hole.
Furthermore, and most interesting for us, the quantum and MONDian conditions together
define the lower (quantum) bound on the mass of a particle (see Fig. 1, note also that $m_{CM} \to 0 $ in the formal limit $h, a_M, 1/c \to 0$). As originally observed by Milgrom \cite{Milgrom1983}, the mass and distance scale estimated in (\ref{mCM},\ref{rCM}) coincide with the hadronic mass scale $m_H$ and the range of inter-hadron interactions (the
proton mass is $m_p=1.67\times 10^{-24}g$, the pion mass is $m_{\pi}=0.25\times 10^{-24}g$). For illustrative purposes we marked the position of $m_p$ in Fig.  1 by the horizontal thin line. 
In 
\cite{ZenMPLA21} it was argued that the conjunction of the quantum/MONDian intersection point with the hadronic scales should not be considered a mere coincidence but that it provides us with arguments in favor of the universal nature of $a_M$ and its relevance for the idea of space emergence. The arguments of \cite{ZenMPLA21} involved the consideration of the basic MOND
relation (\ref{MOND1}) in connection with the peculiar lepton mass formula discovered by Koide \cite{Koide}, both formulas suggesting the possibly fundamental role for the square root of mass. In addition, ref. \cite{ZenMPLA21} mentioned a 
plausible connection of MOND with the internal spatial structure of excited baryons.
Below we will try to shed some additional light on the latter issue.\\

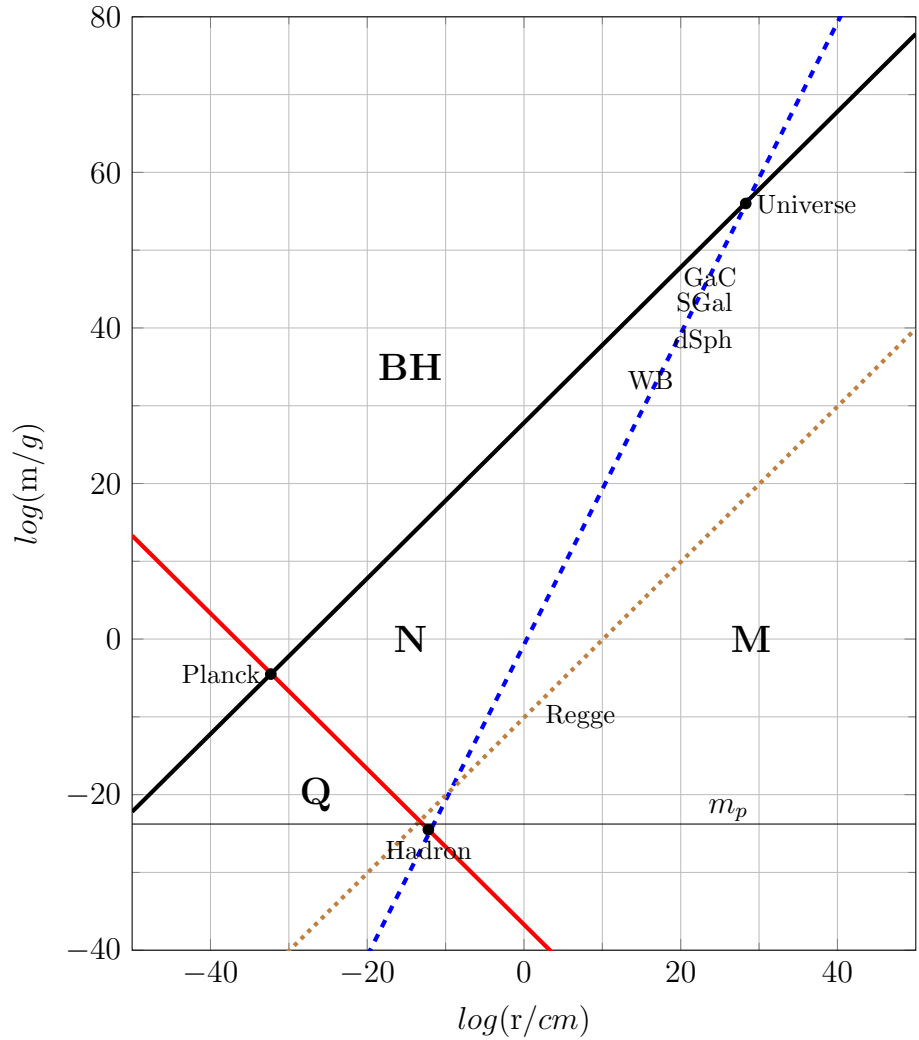
\begin{figure}
\centering
\begin{tikzpicture}
\begin{axis}
[
xmin=-50, xmax=50,
ymin=-40, ymax=80,
xtick={-40,-20,0,20,40},
ytick={-40,-20,0,20,40,60,80},
xlabel={\(log({\rm r}/{cm})\)},
ylabel={\(log({\rm m}/{g})\)},
grid=both,
minor tick num = 1,
width=12.0cm,
height=14.0cm],
\addplot[domain=-50:50,color=black,ultra thick]{x+27.8};
\addplot[domain=-50:50,color=red,ultra thick]{-x-36.7};
\addplot[domain=-50:50,color=blue,ultra thick,dashed]{2*x-0.74};
\addplot[domain=-50:50,color=brown,ultra thick,dotted]{x-10.1};
\addplot[domain=-50:50,color=black]{-23.78};
\filldraw[black] (-32.3,-4.5) circle (2pt) node[anchor=east]{\footnotesize Planck};
\filldraw[black] (28.3,56) circle (2pt) node[anchor=west]{\footnotesize Universe};
\filldraw[black] (-12.2,-24.5) circle (2pt) node[anchor=north]{\footnotesize Hadron};
\filldraw[black] (-18,0) node[anchor=west]{\bf \large N};
\filldraw[black] (-20,35) node[anchor=west]{\bf \large BH};
\filldraw[black] (25,0) node[anchor=west]{\bf \large M};
\filldraw[black] (-30,-20) node[anchor=west]{\bf \large Q};
\filldraw[black] (13,-10) node[anchor=east]{\footnotesize Regge};
\filldraw[black] (30,-22) node[anchor=east]{$m_p$};
\filldraw[black] (28.5,46.6) node[anchor=east]{\footnotesize GaC};
\filldraw[black] (28,43.3) node[anchor=east]{\footnotesize SGal};
\filldraw[black] (28,38.3) node[anchor=east]{\footnotesize dSph};
\filldraw[black] (20.5,33.3) node[anchor=east]{\footnotesize WB};
\end{axis}
\end{tikzpicture}
\caption{{\footnotesize Newtonian Triangle. Thick solid lines define the mass-dependent radii corresponding to Newtonian/black hole (N/BH) and Newtonian/quantum (N/Q) boundaries. Thick dashed line specifies the Newtonian/MONDian (N/M) transition region. Approximate positions of galaxy clusters GaC, spiral galaxies SGal,
dwarf spheroidal galaxies dSph, and wide binaries WB are shown (taken from Fig. 1 in \cite{Hernandez}).
The horizonthal thin line corresponds to proton mass.
The dotted line, calculated from Regge slope $\alpha'$, shows the dependence of hadronic mass on string length.}} \label{M1}
\end{figure}

 \section{Beyond Newtonian Triangle: Excited Hadrons}
In this section we will extend our earlier arguments in favor of the universal and
 fundamental nature of $a_M$. To this aim we will supplement
the equality of $(m_{CM},r_{CM})$ and $(m_H,r_H)$ (quantum/MONDian and hadronic) scales
with yet another
hadron - MOND `coincidence'. 
We start with a brief review of some very general properties of hadronic spectrum, 
which lead to the conclusion that, with their mass growing, the excited hadrons 
intrude more and more into the MONDian regime.  
This observation directs our attention to the internal spatial structure of excited baryons in the quark model, suggesting a
link between the problem of `missing states' in the baryonic spectrum, the MOND approach, and the idea of emergent space.

For our purposes, the most important aspect of hadron's internal structure
is the existence of interquark strings (flux tubes in QCD) that ensure quark confinement.
The resulting picture of excited hadrons is that of string-like states
whose mass is steadily growing when the string becomes longer and longer. 
This string is characterised by a universal constant - the strength of
interquark string tension. Its size may be estimated directly from experiment and, consequently, it is 
independent of the details
of interquark dynamics.
 The relevant feature of the observed hadronic spectrum is the existence of linear `Regge trajectories' which describe infinite `towers' of similar excited states, linking their spin J and mass m via the formula: 

\begin{equation}
\label{Rtraj}
J=\alpha_0+\alpha'm^2,
\end{equation}
where $\alpha_0$ is the so-called intercept and $\alpha'$ -- the slope of the trajectory.
For example, the recurrences of nucleon occur for spins $J=1/2, 5/2, 9/2,... $ with masses given via (\ref{Rtraj}).
The slopes of all trajectories are similar, thus permitting the introduction of
a universal constant $\alpha'$ which was estimated
to be \cite{Zen2018}:
\begin{equation}
\label{alpha'}
\alpha'\approx 0.38 \times 10^{21}cm^2/(g~s).
\end{equation}
From (\ref{alpha'}) one can form the constant
 \begin{equation}
 \label{kappamr}
 \kappa_{m,r} =c/\alpha'=0.8 \times 10^{-10}g/cm,
 \end{equation}
which permits to express the masses of excited hadrons in terms of  the lengths of interquark strings:
\begin{equation}
\label{mR}
m=\kappa_{m,r}~r.
\end{equation}

The related constant $\kappa_R=c~\kappa_{m,r}=2.37 ~g/s$ allows the expression of momenta as
proportional ({\it not inversely proportional}) to positions and permits the consideration of the 6D nonrelativistic (momentum {\it and} position) phase space as a natural generalization of the ordinary 3D (position {\it or} momentum) space. It may be shown \cite{Zen2008,ZenBook,Clifford} that the symmetries of the classical 6D phase space - when considered in conjunction with position-momentum quantum commutation relations - lead to the 
 appearance of internal quantum numbers identifiable with hypercharge and color, and the 
emergence of the familiar $U(1) \otimes SU(3)_C$ symmetry of the Standard Model. In fact, they provide a natural string-related explanation of the 3+1 
quark-lepton pattern of a single generation of elementary fermions, exactly as constructed in the
 Harari-Shupe rishon model \cite{HSmodel}. This supplies further argument that the introduction of the mass-distance conversion constant $\kappa_{m,r}$ of Eqs (\ref{kappamr},\ref{mR}) touches on some fundamental physics. \\

The Regge connection (\ref{mR}) between excited hadron mass $m$ and the length of the corresponding interquark string $r$ is visualised in Fig. 1 as the dotted line. It is parallel to the
black hole side of the Newtonian triangle because of the equality of dimensions of the Regge ($\kappa_{m,r}= 0.8 \times 10^{-10}~g/cm$) and black hole ($c^2/2G = 0.67 \times 10^{28}~g/cm$) mass-length conversion constants (note that the two constants differ by a factor of order $10^{40}$). 
Note also that up to a factor of 5 or 10 the Regge line crosses the quantum side of the triangle at the proton mass. The coincidence of the MONDian/quantum and Regge/quantum crossing points is not very strange. In fact, it is expected as the spacing of Regge recurrences is of the order of proton mass squared.  The important point about the Regge line is that - for larger masses and string lengths - it points farther and farther into the MONDian regime.  Thus, MONDian universality assumed, the properties of excited hadrons could provide us with important information on MOND in the quantum particle region. In this respect, excited baryons turn out to be particularly interesting. Below we present a highly simplified account
of a crucial aspect of their spectroscopy.

Baryon spectroscopy is usually discussed with the harmonic-oscillator 
nonrelativistic constituent quark model (NR CQM) in mind as a benchmark for more sophisticated approaches. 
In this model the
excited baryons are constructed as three-quark states with spatial/orbital excitations 
in the two a priori possible inter-quark spatial degrees of freedom. Experimental limitations restrict
the range of observable excited baryons mainly to states build from the light quarks (flavors $u,d,s$).
With the Pauli-principle-dictated antisymmetry of the three-quark 
wave function
ensured by color degree of freedom, and with quarks of definite flavor and spin ($u\uparrow$,~$u\downarrow$,~$d\uparrow$,~...,~$s\downarrow$) classified in the fundamental
representation  {\bf 6} of $SU(6)$ symmetry group, the expected spectrum of baryons 
 coresponds to  totally symmetric (in spin-flavor-space)  wave functions that may be constructed for three quarks in various configurations (classified in $SU(6) \otimes O(3)$
 supermultiplets, with $O(3)$ describing the orbital motion).
 
 In general, the system of baryonic quarks may belong to  one of the $SU(6)$
 multiplets obtained from the product of three ${\bf 6}$'s:
 \begin{equation}
 \label{666}
 {\bf 6}\otimes {\bf 6} \otimes {\bf 6} = ({\bf 21}_S \oplus {\bf 15}_A) \otimes {\bf 6} =
 ({\bf 56}_S \oplus {\bf 70}_M) \oplus ({\bf 70}_M \oplus {\bf 20}_A),
 \end{equation}
 where indices S,M,A (symmetric, mixed, antisymetric) were added to indicate spin-flavor symmetry of the multiplet in question.
  As the spin-flavor-space w. f. has to be totally symmetric, the ground-state (spatially symmetric) baryons must belong to the ${\bf 56}$ (this includes nucleon in particular). The existence of the remaining spin-flavor multiplets of (\ref{666}) requires 
that the total symmetry of the overall w.f. be restored by appropriate spatial/orbital inter-quark excitations. This conclusion refers in particular to the antisymmetric `diquark' ${\bf 15}$ which is essential for building the ${\bf 20}$. Yet, when one identifies the experimentally observed states with the members of $SU(6)$ multiplets, the ${\bf 20}$ multiplet does not appear. This suggests the lack of appropriate spatial/orbital
excitations `inside' the antisymmetric diquark ${\bf 15}$ and the absence of ${\bf 15}$ in general
(as well as strong quark-quark clustering into a symmetric diquark ${\bf 21}$). In effect, many states should be absent from the expected (in NR CQM) spectrum
of excited baryons, a situation that experimentally seems to be qualitatively confirmed. This is the ``missing baryons'' problem of elementary particle physics.\\

There are two possible solutions to this
problem. According to the first solution, for some reasons
the number of internal spatial degrees of freedom is strongly reduced. In particular,
one might think that the NR CQM description is highly oversimplified and misses some dynamical QCD effects that are basically equivalent to the appearance of spin-flavor symmetric diquarks and the absence of spin-flavor antisymmetric diquarks. However, lattice QCD calculations \cite{MissingBaryonsLattice2011} 
 yield a spectrum of states identifiable as admixtures of $SU(6) \otimes O(3)$ 
 representations and the counting of levels that are consistent with the NR CQM. In other words, the calculated spectrum is incompatible with the quark-diquark solution to the problem.
According to the second solution, the missing states are not observed because 
they couple very weakly to decay channels used in the experimental analyses, thus escaping detection.\\

With this situation in mind, in their turn-of-the-century review of baryon spectroscopy Capstick and Roberts wrote \cite{Capstick}:
  ``These questions about baryon physics are fundamental. If no new baryons are found, both
QCD and the quark model will have made incorrect predictions, and it would be necessary
to correct the misconceptions that led to these predictions. Current understanding of QCD
would have to be modified, and the dynamics within the quark model would have to be
changed." Today, after 20 years, the problem is still there, with ``only a small percentage of the predicted states (...) found."\cite{MissingBaryons2019}\\
 
 \section{Conclusions}
In this situation I believe it is time to take seriously the possibility that one of the two a priori existing internal spatial degrees of (quark) freedom is completely frozen or even non-existent. With the highly excited hadrons clearly belonging to the MOND regime (Fig. 1), the empirical observation of diquark clustering in excited baryons seems to tell us something important about MOND itself, and, in particular, about
its connection to space. As a matter of fact, freezing or irrelevance of one spatial degree of freedom could be anticipated directly from the basic MOND formula (\ref{MOND1}). 
Indeed, as pointed out in \cite{ZenMPLA21}, the proportionality of $a$ to $1/r$ suggests that
in the MOND regime the lines of gravitational field are squeezed or restricted to two dimensions only (just as the proportionality of Newtonian force to $1/r^2$ is associated
with the existence of three dimensions). If we take this option seriously - and the spectrum of excited baryons suggests just that - this may be a hint
that the familiar 3D space is somehow constructed from its 2D subspaces \cite{ZenMPLA21}, and that the spectrum of elementary particles (hadrons in particular) holds important information on the quantum
aspects of the emergence of space, as originally argued in \cite{Zen2018}.\\

\vfill

\vfill

\end{document}